\documentclass[conference]{IEEEtran}
\IEEEoverridecommandlockouts
\usepackage{cite}
\usepackage{amsmath,amssymb,amsfonts,nccmath,tabularx,booktabs}
\usepackage{graphicx}
\usepackage{float}
\usepackage{subfigure}
\usepackage{textcomp}
\usepackage{hyperref}
\usepackage{multicol}% http://ctan.org/pkg/multicols
\usepackage{graphicx}
\usepackage{xcolor}
\usepackage{svg}
\usepackage{algorithm}
\usepackage{algpseudocode}
\def\BibTeX{{\rm B\kern-.05em{\sc i\kern-.025em b}\kern-.08em
    T\kern-.1667em\lower.7ex\hbox{E}\kern-.125emX}}
\begin{document}

\title{RIDS : Real-time Intrusion Detection System for WPA3 enabled Enterprise Networks \\

% \thanks{We are grateful of Arista Networks for funding this reasearch.}
}

\author{
\IEEEauthorblockN{Rahul Saini, Debajyoti Halder, and Anand M. Baswade}
\IEEEauthorblockA{Dept. of Electrical Engineering and Computer Science \\
Indian Institute of Technology Bhilai, India \\
Email: \{rahuls, debajyotih, anand\}@iitbhilai.ac.in}
}

\maketitle

\begin{abstract}
With the advent of new IEEE 802.11ax (WiFi 6) devices, enabling security is a priority. Since previous versions were found to have security vulnerabilities, to fix the most common security flaws, the WiFi Protected Access 3 (WPA3) got introduced. Although WPA3 is an improvement over its predecessor in terms of security, recently it was found that WPA3 has a few security vulnerabilities as well. In this paper, we have mentioned the previously known vulnerabilities in WPA3 and WPA2. In addition to that, we have created our own dataset based on WPA3 attacks (Section \ref{attacks}). We have proposed a two-stage solution for the detection of an intrusion in the network. The two-stage approach will help ease computational processing burden of an AP and WLAN Controller. First, AP will perform a lightweight simple operation for some duration (say 500ms) at certain time interval. Upon discovering any abnormality in the flow of traffic an ML-based solution at the controller will detect the type of attack. Our approach is to utilize resources on AP as well as the back-end controller with certain level of optimization. We have achieved over 99\% accuracy in attack detection using an ML-based solution. We have also publicly provided our code and dataset for the open-source research community, so that it can contribute for future research work.
\end{abstract}


\section{Introduction}
Since the inception of WiFi in 1997, it has been widely successful as a de facto wireless technology. Today we have billions of WiFi devices around the world. It has reached everywhere from industries to business houses to the common households. It is a gateway to access the internet. Although accessing the internet is cardinal these days, the security of these devices is also very important. 

Over the years WiFi has improved its security features. The latest IEEE 802.11ax (WiFi 6) is an improvement over its predecessor IEEE 802.11ac (WiFi 5) in the context of security. The introduction of Wireless Protected Access 3 (WPA3) in WiFi 6 has eliminated many previously known attacks in WPA2 accessible devices. WPA3 mandates the use of dragonfly handshake, which is based on Simultaneous Authentication Equals (SAE). It provides encryption to open wireless Access Points (AP) as well, which is  advantageous with respect to security.

WiFi 6 came with security enhancements but with a few security vulnerabilities as well. It has been found that WPA3 is vulnerable to attacks categorized as denial-of-service, side-channel, and downgrade attacks. WiFi devices need to be backward compatible, in order to serve the users utilizing previous generation stations. To do this WiFi 6 has to downgrade itself for the user stations compatible with WPA2. This compromised the security of WPA3 with WPA2 attacks. Some of the most popular penetration testing tools like Metasploit, Aircrack-ng, and MDK3 are being IEEE widely used for testing attacks against wireless devices. All these tools are publicly available and have been used by network researchers. Since IEEE 802.11ax devices are not widely in use, security research on these devices are still being explored. 

To mitigate this, datasets are used to study the vulnerabilities in the WiFi protocols. AWID3 is one such dataset \cite{awid3} which has been used in multiple researches of Intrusion Detection in WLAN including deterministic as well as ML-based approaches \cite{awid3paper3, awid3paper1, arista, awid3paper2}. AWID3 dataset is a collection of PCAP and CSV files of various intrusions into a WLAN test-bed. However this dataset is limited to WPA2 security protocol only. 
Development of detection mechanism over the attacks on wireless devices is the primary focus of the paper. Since there is no WPA3 dataset publicly available we created our own dataset by performing attacks on our WLAN test-bed. Our code is available on GitHub \cite{code-rids}. 

This research focuses on the shortcomings of WPA3 and suggests a few techniques for intrusion detection. We propose a Real-time Intrusion Detection System (RIDS) for enterprises with WLAN to detect an intrusion into their network in real-time. The primary contributions of the paper are:
\begin{enumerate}
    % \item Access points (APs) are resource-constrained when it comes to performing intense computational tasks. Still, APs can play a vital role in determining whether an attack is being performed on the network.
    \item A two stage architecture for real-time intrusion detection for an enterprise scenario. The stages comprise of an Intrusion Alert System to raise an alert if an intrusion is detected, and an ML-based classifier to predict the attack type and raising a network-wide alarm.
    \item Created a new dataset for attacks on WPA3-enabled device and have made it publicly available for further research.
    \item A lightweight ML-based classifier for attack detection with high accuracy in real-time. We propose a Random Forest Classifier trained on our dataset for predicting attacks.
\end{enumerate}

\section{Related Works}
WPA3 got introduced in 2018 as a global WiFi standard~\cite{wpa3-cert}. WiFi devices operating on WPA2 are not secure \cite{arista, wpa2weak, wpa2comm} and are vulnerable to attacks \cite{wpa2attacks, dragonblood}. WPA3 had some convincing improvements in regarding the security of the handshake protocol. However it has also been prone to attacks as well \cite{wpa3weak, krack}. A lot of work has been done on an intrusion detection system (IDS) and different methods have been introduced by many for effective detection \cite{ids1, ids2, ids3, dragonshield}. The goal has always been to detect an attack, however there haven't been any lightweight and cost-efficient solution that is feasible enough to detect in real-time an ongoing attack.  

Intrusion detection systems have not been limited to deterministic algorithms only. ML-based classifiers have also been implemented to predict attacks from captured sequence of frames \cite{ml1, ml2, ml3}. ML-based solutions have shown better accuracy in predicting an attack, however the classifiers could not have been used to detect an ongoing attack. IDS on the other hand is slow in detecting an attack using the deterministic algorithm. Therefore we require a different approach for the detection phase of an IDS to detect an ongoing attack. We planned to use ML-based classifier to get an accurate prediction. The AWID3 dataset based on WPA2 has provided researchers with a platform to work on ML-based intrusion detection classifiers. However, the dataset is completely based on WPA2 attacks and most of the attacks mentioned are not valid in WPA3. To tackle this problem, we have created our dataset and presented the findings.

We have a two-way approach to club together with the benefits of both signature-based IDS and machine learning for attack detection, where IDS gets triggered on attack suspicion and ML solution is for detection of an attack. This paper presents the proposed solution to our approach. We have presented our code and files \cite{code-rids}. This may help other interested researchers to have a better understanding of WiFi attacks and their detection.

\section{Attacks on WPA3}
\label{attacks}

\begin{figure}[h]
    \centering
    \includegraphics[width=\linewidth]{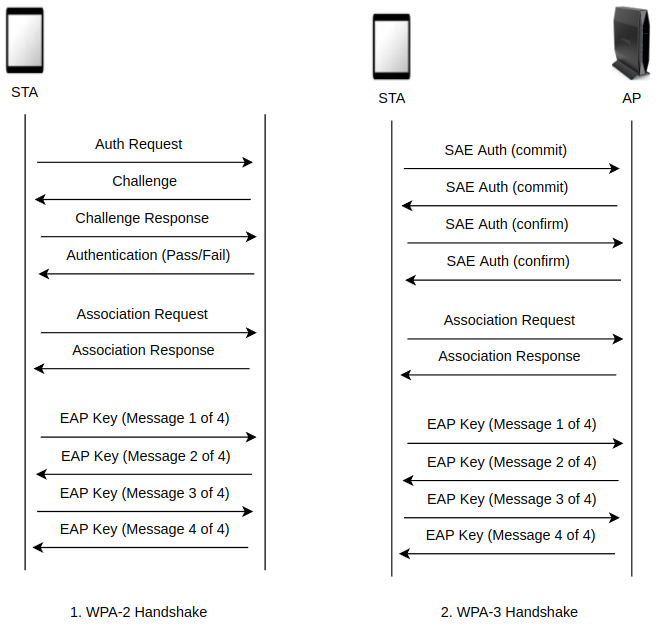}
    \caption{(1) WPA-2 (Shared Key Authentication), (2) WPA-3 (Simultaneous Authentication Equals)}
    \label{fig:handshake}
\end{figure}

We performed the attacks in our test-bed (explained in Section \ref{setup}) and captured the packets for training our ML model. Figure \ref{fig:handshake} shows the abstract representation of message exchange between an AP and STA. After authentication and association messages, 4-way handshake using Extensible Authentication Protocol Over LAN (EAPOL) takes place \cite{arista}. The third message of 4-way handshake containing Robust Security Network Element (RSNE) provides the information about the protocols and parameters supported by the AP. In case of any discrepancy the connection is aborted. 
The attacks in focus are as follow:
\subsection{De-authentication}
The attack is possible in WPA2 and WPA3 as well. An attacker can force the STA to disconnect from AP using a de-authentication attack. WPA3 refrains attackers from spoofing the de-authentication frame once a 4-way handshake is completed, due to Management Frame Protection (MFP). Although an attacker can still de-auth the STA by sending multiple de-authentication frames right after the association request from an STA to an AP. Thus results in creating an ambiguity for the STA to process succeeding packets from the AP. This will result in STA sending a de-authentication packet to the AP. Thereby, the STA will disconnect from AP.

% \subsection{WPA3 Transient Mode}
% WiFi devices are supposed to be backward compatible. This allows users with the previous version of devices to connect with the latest APs. The transition state is vulnerable to attacks. An attacker can configure hostapd 2.9 in such a way that it can transmit beacons at a higher rate, advertising a WPA2 connection for legitimate AP.  This will allow a victim device to connect to an AP with a WPA2 connection instead of a WPA3. The victim while the exchange of the third message of the handshake between STA and an AP. When AP will realize the STA is connecting using a WPA2 connection, will allow the STA to establish the connection using a WPA2 connection. This would expose the AP to WPA2 attacks.

\subsection{Rogue AP}
This attack involves configuring a fake AP using hostapd. Once a fake AP is created, an attacker can launch the AP with the same name, bssid as a legitimate AP, and broadcast at a lower beacon interval. This will let the STA trust the rogue AP and try to connect with it. In the case of the WPA3 AP, STA can be spoofed to connect with the WPA2 connection through our rogue AP. By lowering the beacon interval rogue AP can broadcast a WPA2 connection instead of WPA3. This will make STA connect using WPA2 authentication, STA will later acknowledge in message 3 of the 4-way handshake that AP connects using WPA3 authentication. This will result in denial of service for the STA, as it would detect the connection is based on WPA3 instead of WPA2.

\subsection{Evil Twin}
The attack is based on creating a fake AP. Just like we did in rogue AP. This time attacker tries to disconnect the STA from legitimate AP and connect to Evil Twin AP (ETAP). This will work in a scenario of open AP present in public places. If the AP is password protected and runs on the WPA2 authentication protocol then an attacker can run an offline dictionary attack on the AP. Once the password is discovered, an attacker can create ETAP entirely the same as legitimate AP.  This would result in a man-in-the-middle attack, flooding the network with garbage data, or even denial of service.

\subsection{Krack}
The Key Reinstallation attack manipulates the 4-way handshake in the WPA2 protocol. It allows the attacker to block messages 2 or 4 of the WPA2 handshake, which in turn results in the retransmission of messages 1 or 3.  It takes the advantage of AP retransmitting the same message to STA. This would result in the installation of the same PTK (Pairwise transient key) or GTK (Group transient key) on STA. However, the latest windows, android, and iOS don’t accept retransmission of message 3 of the 4-way handshake. Thus remain secure against the Krack attack. For our test-bed, we first conducted the Rogue AP attack to downgrade the connection to WPA2, and then performed the Krack attack.

\subsection{Beacon Flooding}
The attack is simple and does not require any intervention between STA and an AP. The beacon frames are not protected. An attacker can easily exploit them by flooding multiple beacon frames compared to a genuine AP. Sending multiple probe responses would increase the computation time for the STA to find the legitimate AP. If the probe responses are similar to an AP, then this would end up in denial of service to the STA. Since STA would try to connect to an Service Set Identifier (SSID) with similar identification.

\section{Experimental Setup}
\label{setup}

\begin{figure}[h]
    \centering
    \includegraphics[width=\linewidth]{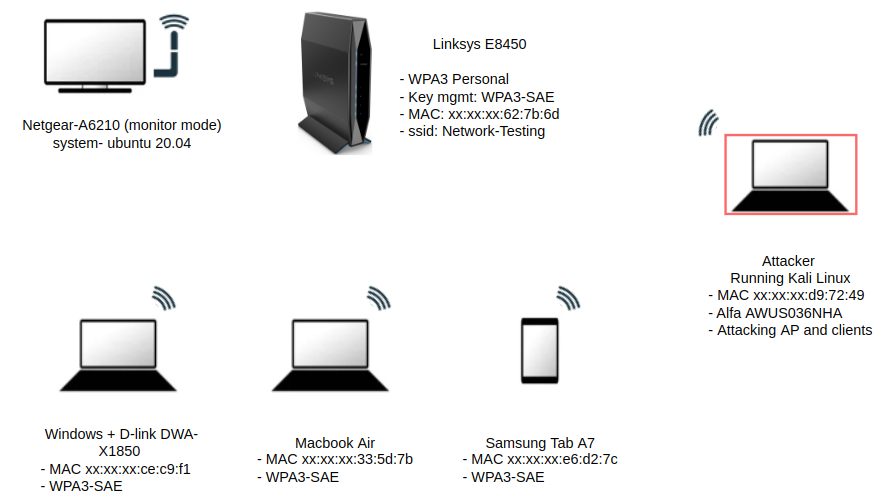}
    \caption{Experimental Setup.}
    \label{fig:setup}
\end{figure}

To perform the attacks, we set up our own test-bed setup. We created a dataset of attack vectors which is generic and can be easily expanded. Our dataset is created using different devices as shown in \ref{fig:setup}. For WPA3 enabled access point, we used the Linksys E8450 device and for the WPA3 WiFi adapter, D-link DWA-X1850 was used. One Alfa AWUS036NHA adapter (Atheros AR9271 chipset) was used for monitoring the channel and injecting packets for the purpose of performing attacks. Netgear A6210 device is used for monitoring traffic between AP and STAs. Netgear device was connected to a desktop running Ubuntu 20.04. For STAs we have used a Samsung A7 tablet, MacBook Air, and an HP laptop running windows 10 using a D-link adapter supporting WPA3. 

We used Linksys AP which supports IEEE 802.11ax and it is running in WPA3 mode on a 2.4 GHz frequency. All our attacks are performed on 2.4 GHz frequency only. The 5GHz frequency was also working but no attacks were performed on the 5GHz frequency. The Netgear A6210 adapter was used for the purpose of capturing packets. Figure~\ref{fig:setup} is just a representation of our setup for testing and collecting data. 

Initially, we assumed that the WPA3 connection mandates the usage of MFP. However, in our experiment, we have found that when AP and STA both are WPA3 compatible, we were able to de-authenticate the client simply by flooding de-auth frames. We have performed all our experiments without manually switching on the MFP. This was done to examine if MFP is used automatically or not.

To create the dataset, we had to label each frame if it was responsible for a particular attack. This initial detection model is based on traffic analysis of certain frames. The mechanism is primarily based on the following frames:
\begin{enumerate}
    \item Beacon Frame
    \item Authentication Frame
    \item De-authentication Frame
    \item Association Frame
    \item Dis-association Frame
    \item EAPOL Frame
\end{enumerate}

The resulting dataset was a collection of packet captures constructed from multiple attack sessions with a total of 250 attributes. The attacks considered in this research are De-authentication, Rogue AP, Beacon Flooding, Evil Twin and Krack attacks. The dataset has CSV files that contain the packets transmitted in the network while the attacks were being performed. These packets can be analysed using a deterministic algorithm (as shown in Section \ref{attacks}) to detect the attacks and also using ML to find a correlation between the attributes to detect the attacks. To the best of our knowledge, this is the first dataset of WPA3 attacks that have been created out of real-life intrusion experiments on a test-bed.

\section{Proposed Real-time Intrusion Detection System}
\label{rtids}

Enterprises have WLAN architecture with a controller communicating with the access points to monitor and manage the traffic in the network. Controllers can be used to periodically monitor and analyse the incoming packets to detect any attacks on the network. However, analysing each and every packet in the network for probable WiFi attacks is not feasible since it is very computation-intensive for one controller and also consumes network bandwidth between AP and Controller. The delay in detection is crucial for enterprises with massive networks since a downgrade attack generally takes a few seconds to make a device connect to a rogue AP. In that time if the controller queue is busy analysing packets from another AP then the attack may never be detected. 

\begin{figure}[h]
    \centering
    \includegraphics[width=\linewidth]{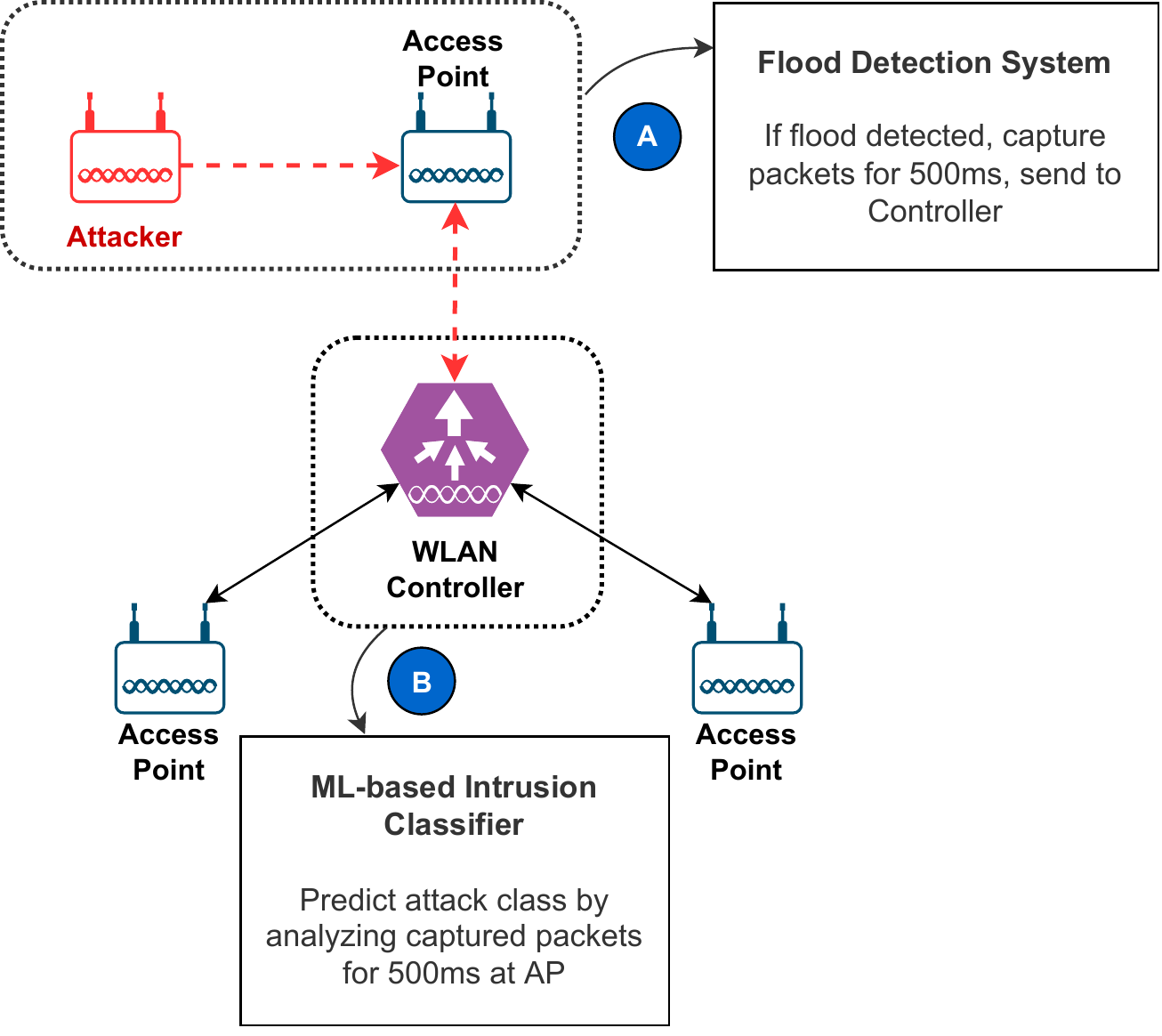}
    \caption{Architecture for Attack Detection System.}
    \label{fig:ads_arch}
\end{figure}

We propose a Real-time Intrusion Detection System (RIDS) comprising of 2 checkpoints of detection. The first checkpoint is at the AP, where we run our Flood Detection System (FDS) to detect if there is a sudden flood of packets at the AP. If a sudden flood is detected (which might be even caused by a genuine request) the AP captures all the packets received for the next 500ms. The attacker would have to send packet at lower time interval to make sure the success rate of an attack \cite{dragonblood}. Keeping the time interval at 500ms, which is higher and provides balance between unnecessary over and under detection~\cite{arista}. Once any anomaly detected, sends the packets to the controller. At the second checkpoint in the controller, the controller passes the captured packets through an ML-based Intrusion Classifier. This Intrusion Classifier has been trained on WPA-3 and WPA-2 intrusion packets to predict the intrusion type (if any) with 99\% accuracy. If the classifier predicts an intrusion then we can definitively say that an intrusion has occurred. The network is alarmed about the intrusion. However, the controller continues analysis on the captured packets, to find the source device which initiated the attack. This information can be flooded in the network so that all the APs can block packets coming from this device.
Fig. \ref{fig:ads_arch} shows the architecture for the Real-time Intrusion Detection System and the steps taken by the system when an attacker tries an intrusion into the network. 

\begin{figure}[h]
    \centering
    \includegraphics[width=\linewidth]{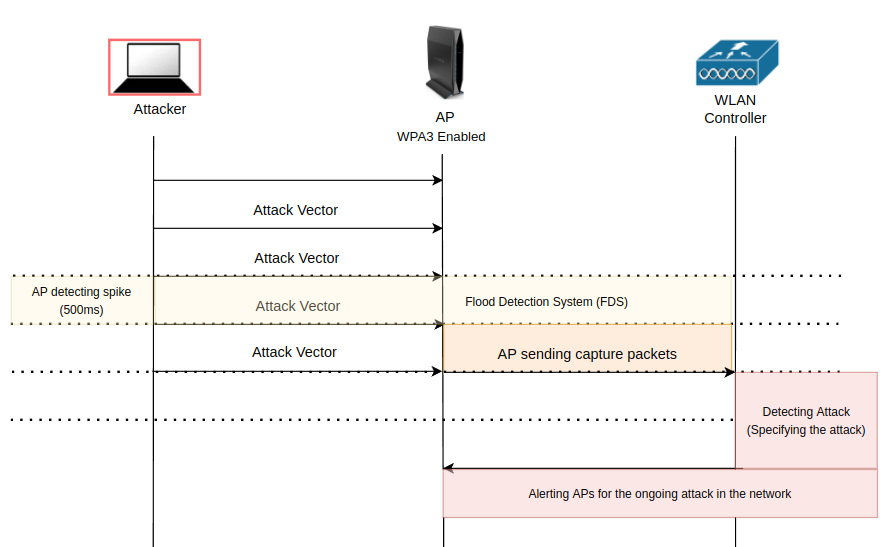}
    \caption{Flow diagram for Attack Detection System.}
    \label{fig:atk_flow}
\end{figure}

If the controller had been flooded with all the packets received at all the APs then it would be impossible to detect an intrusion in real-time as there would be long queues of packets waiting to be analyzed. RIDS distributes the initial detection phase to the APs. With the help of FDS we are alerted whenever there is a chance of intrusion. To detect a flood of packets, using ML would not be an optimal solution as floods can be easily identified by analyzing the interval between the packets. However to detect an attack with high accuracy, using ML turned out to be the best solution. ML classifiers are lightweight (our ML-based Intrusion Classifier is just 1.7MB in size) and are fast to make predictions. They can be used for real-time prediction jobs with high accuracy. Therefore, the WLAN Controller in equipped with an ML-based Intrusion Classifier which analyzes packets received from intruded APs to detect an on-going attack. Fig \ref{fig:atk_flow} presents an abstract view of flow control between an AP and WLAN controller while an attack being on going in the network.

\subsection{Flood Detection System (FDS)}

At every AP, we run our FDS Algorithm~\ref{Algo:Flood} to detect if a flooding has been conducted at a particular AP. For every 1 sec time quantum, we find the number of incoming packets by calculating the difference between the frame numbers at the beginning and at the end of the time quantum. Intuitively, a spike in the number of incoming packets might infer a flood. However, the spike might be caused by genuine traffic from a high number of users. Therefore we find the mean number of packets per user connected to the AP. If the mean has spiked by more than 10 times then we classify it as a flood. In case of a DDoS attack, the mean might not spike more than 10 times. In such cases, if the total number of packets spike by more than 15 times, we consider it as a flood. 

\begin{algorithm}[!htb]
\caption{Flood Detection Algorithm}
\label{Algo:Flood}
\begin{algorithmic}[1]

\Procedure{Detect Flood At AP}{$ap\_id$}
    \Procedure{Interval}{1000}
        \State diff $\leftarrow$ abs(frame\_in - frame\_out)
        \State mean\_diff $\leftarrow$ diff / users
        \If{mean\_diff $>$ old\_mean\_diff $\times$ 10}
            \State pkt $\leftarrow$ capture(500, ap\_id) \Comment{capture incoming packets for next 500ms}
            \State sendToController(pkt, ap\_id)
        \Else \If{diff $>$ old\_diff $\times$ 15}
                \State pkt $\leftarrow$ capture(500, ap\_id)
                \State sendToController(pkt, ap\_id)
            \EndIf
        \EndIf
    \EndProcedure
\EndProcedure

\end{algorithmic}
\end{algorithm}

\subsection{ML-based Intrusion Classifier}

If FDS detects a flood in the network, then we capture the incoming packets for the next 500ms and send them to the Controller (Section \ref{rtids}). These packets are then passed through the ML-based Intrusion Classifier to predict the attack class. We used our dataset to train a few ML classifiers. We tested three different classifiers: 

\begin{enumerate}
    \item Logistic Regression
    \item Decision Tree
    \item Random Forest
\end{enumerate}

\begin{figure}[h]
    \centering
    \includegraphics[width=\linewidth]{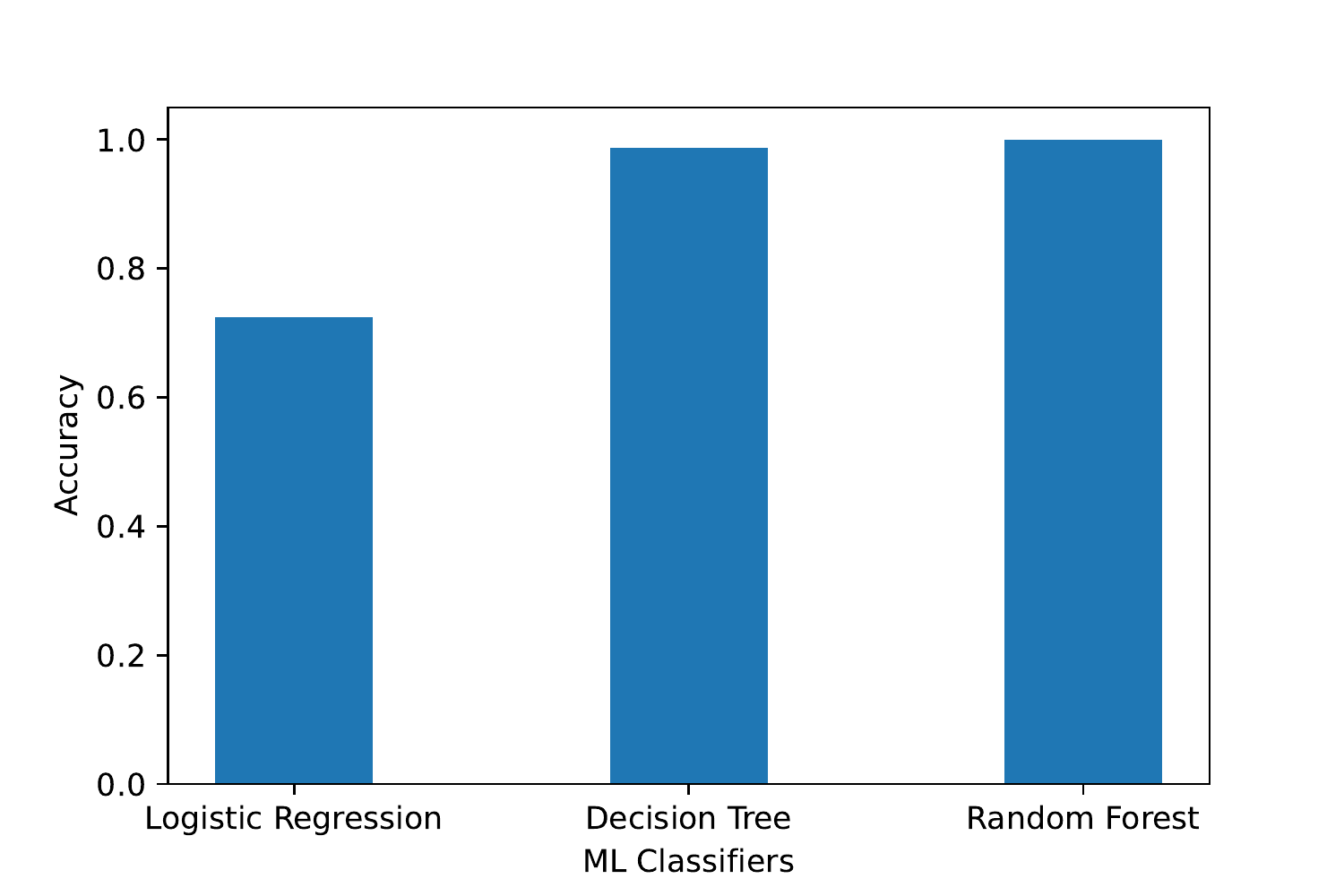}
    \caption{Accuracy by Classifiers.}
    \label{fig:accuracy}
\end{figure}

\begin{figure}[h]
    \centering
    \includegraphics[width=\linewidth]{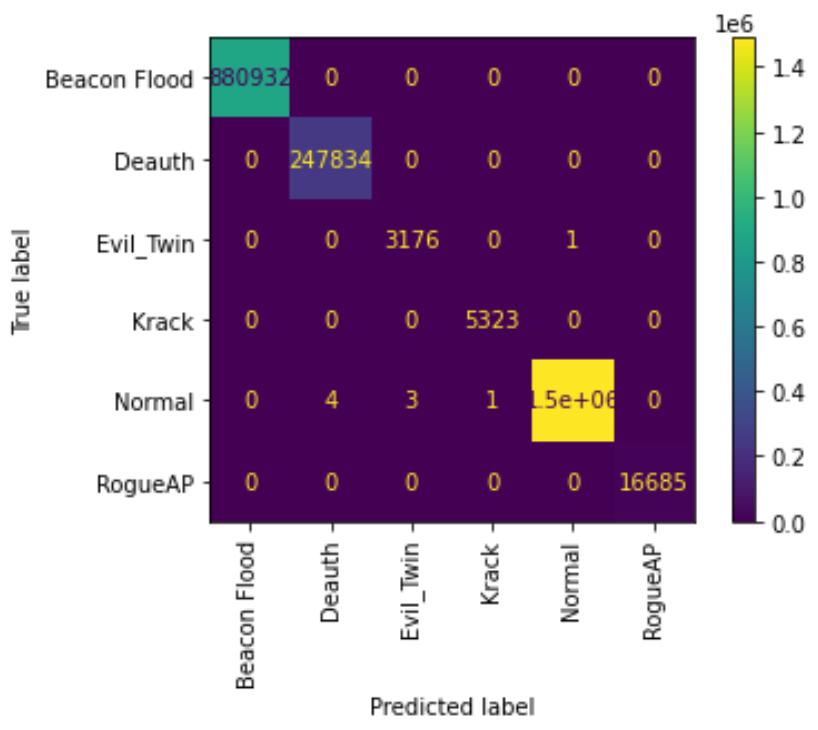}
    \caption{Confusion Matrix for Decision Tree.}
    \label{fig:dt_cm}
\end{figure}

\begin{figure}[h]
    \centering
    \includegraphics[width=\linewidth]{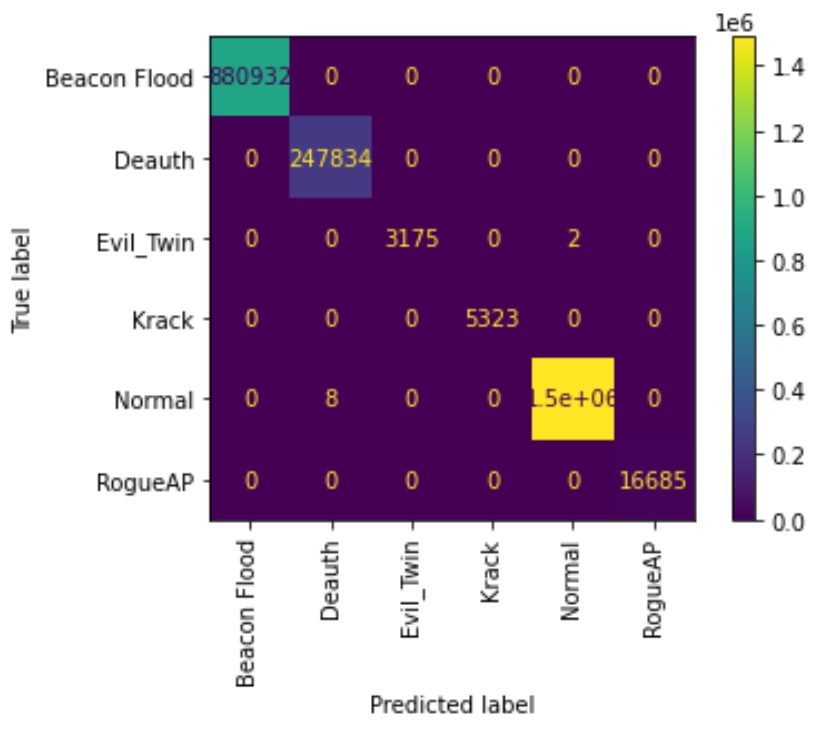}
    \caption{Confusion Matrix for Random Forest.}
    \label{fig:rf_cm}
\end{figure}

\noindent
As shown in Fig. \ref{fig:accuracy}, the accuracy of Decision Tree (99.98\%) and Random Forest (99.97\%) are comparable. We need to remember that the correct evaluation of a classifiers cannot be done just by comparing the accuracy. So we tested the classifier with multiple attack vectors and computed the confusion matrix for each classifier. As seen in the confusion matrices (Figs. \ref{fig:dt_cm} and \ref{fig:rf_cm}), Decision Tree Classifier has the least number of false positives. 

Decision Tree Classifier, as evident from Table \ref{table:fpr_tpr} has a really low False Positive Rate (FPR). We decided to choose Decision Tree Classifier to predict the occurrence of an attack in real-time. Once FDS rings an alert then the Classifier can start analysing packets and detect if the network has been attacked. This mix of algorithms (FDS and ML-based Intrusion Classifier) is useful for a network with very high bandwidth and multiple APs such that monitoring the network centrally for vulnerabilities is not feasible. With such a low false positive rate, Decision Tree Classifier can precisely detect if the network is being attacked upon being triggered by the FDS algorithm. 

\begin{table}[h]
\centering
\caption{Comparison of FPR and TPR of Classifiers}
\label{table:fpr_tpr}
\begin{tabular}{|l|l|l|}
\hline
Metric & Decision Tree & Random Forest  \\ \hline \hline
FPR    & 0.00054        & 0.00068       \\ \hline
TPR    & 0.99946        & 0.99932       \\ \hline
\end{tabular}
\end{table}

\section{Conclusion and Future Scope}

The ML-based Intrusion Classifier predicts the attack type with very high accuracy. Based on the predicted attack type, the controller can continue analyzing the packets to find which devices that are connected to the AP are causing the attack. This information can be shared in the network with all the APs so that APs can take precautions. RIDS as a complete system is apt for securing an enterprise WLAN from an intrusion in real-time. Our goal was to propose a system that is quick enough to raise an initial alert. This is achieved by the FDS which detects a spike in incoming packets almost instantly. After an initial alert, it is required to make sure if an intrusion is being conducted. RIDS captures all the incoming packets for the next 500ms (Section \ref{rtids}). These packets are analyzed by the Controller to find the attack type using an ML-based Classifier. As the Classifier is lightweight, analyzing 500ms packets is instantaneous as well. Even if there are parallel attacks in the network, since the initial FDS is distributed in the APs, the controller is not bombarded with all the incoming attack packets at once. 

RIDS is lightweight and cost-efficient enough to be implemented in an enterprise scenario. As a future work, We plan to implement other attacks and create an even bigger dataset for the number of attacks on WPA2/3. Our plans are in line with WiFi security and to promote secure usage of wireless LAN devices.

\bibliographystyle{./bibliography/IEEEtran}
\bibliography{main}

\vspace{12pt}

\end{document}